\documentclass{PoS}

\usepackage{amsmath}

\newcommand{\be}{\begin{equation}}
\newcommand{\ee}{\end{equation}}
\newcommand{\bea}{\begin{eqnarray}}
\newcommand{\eea}{\end{eqnarray}}
\newcommand{\bt}{\begin{tabbing}}
\newcommand{\et}{\end{tabbing}}
\newcommand{\bi}{\begin{itemize}}
\newcommand{\ei}{\end{itemize}}
\newcommand{\ben}{\begin{enumerate}}
\newcommand{\een}{\end{enumerate}}

\newcommand{\calO}{{\mathcal O}}

\newcommand{\Dt}{\Delta t}
\newcommand{\Dtp}{\Delta t^\prime}

\newcommand{\NssN}{\left\langle N|\bar ss|N\right\rangle}
\newcommand{\NssNbare}{\langle N|\bar ss|N\rangle_{\rm bare}}
\newcommand{\NllN}{\left\langle N| \bar{u}u + \bar{d}d | N\right\rangle_{\rm bare}}

\title{
\begin{picture}(0,0)(0,0)%
\put(370,50){\makebox(0,0)[l]{\textnormal{\normalsize KEK-CP-250}}}%
\end{picture}%
Lattice studies of hadron physics with disconnected quark loops
}

\ShortTitle{
Lattice studies of hadron physics with disconnected quark loops
}

\author{
Takashi Kaneko for JLQCD Collaboration 
\\
KEK Theory Center, 
High Energy Accelerator Research Organization (KEK),
Tsukuba, Ibaraki 305-0801, Japan 
\\ 
School of High Energy Accelerator Science,
The Graduate University for Advanced Studies (Sokendai),
Tsukuba, Ibaraki 305-0801, Japan
\\
E-mail: \email{takashi.kaneko@kek.jp}
}

\abstract{
Disconnected diagrams give crucial contributions 
to the physics of flavor singlet hadrons 
and to scalar form factors of non-singlet hadrons. 
Naive lattice calculation of the disconnected diagrams,
however, requires a huge number of fermion matrix inversions
and hence a prohibitively large computational cost.
In this article, 
we present recent studies of the flavor-singelt meson spectrum
and nucleon strange quark content
using the all-to-all propagator to calculate 
the relevant disconnected diagrams.
}

\FullConference{
35th International Conference of High Energy Physics - ICHEP2010,\\
July 22-28, 2010\\
Paris France
}

\begin{document}

\section{Simulation method}

Disconnected diagrams give crucial contributions 
to interesting physics including the flavor-singlet meson spectrum
and nucleon strange quark content.
Interesting nature of the former, 
such as the famous $U(1)$ problem and the ideal mixing of vector mesons,
has to be confirmed from first principles,
and the latter is an important parameter
in direct experimental searches for the dark matter.
In this article,
we present studies by the JLQCD Collaboration 
\cite{FlvSngl:Nf3:JLQCD,StrCnt:Nf2:JLQCD:direct,StrCnt:Nf3:JLQCD:direct}
that use a recently proposed method,
namely the all-to-all quark propagator \cite{A2A}.



Our gauge configurations of three-flavor QCD 
are generated at a lattice spacing $a\!\simeq\!0.11$~fm 
using the overlap quark action,
which exactly preserves chiral symmetry on the lattice.
We simulate four values of degenerate up and down quark masses $m_l$,
which cover a region of the pion mass 
$300 \!\lesssim\! M_\pi[\mbox{GeV}] \!\lesssim\! 550$~MeV,
and two strange quark masses $m_s\!\simeq\!m_{s,\rm phys}$ and 
$(5/4) m_{s,\rm phys}$ with $m_{s,\rm phys}$ the physical mass.
The flavor-singlet meson spectrum is calculated 
on a spatial lattice volume $(1.8~\mbox{fm})^3$, 
and we also simulate a larger volume $(2.7~\mbox{fm})^3$ 
for the nucleon strange quark content 
to avoid finite volume effects.
%
We refer readers to Ref.\,\cite{Nf3:Prod_Run:JLQCD} 
for further details of our gauge ensembles.


It is conventional to calculate connected hadron correlators 
by using the so-called point-to-all quark propagator $S_F(x^\prime,x)$,
which flows from a fixed lattice site $x$ to any site $x^\prime$
and is obtained by solving
\bea
   \sum_{x^\prime}D(y,x^\prime)S_F(x^\prime,x)
   & =&
   \delta_{y,x},
   \label{eqn:meas:lineq}
\eea
where $D$ is the Dirac operator.
One can naively evaluate disconnected quark loops $S_F(x,x)$
by solving (\ref{eqn:meas:lineq}) for each lattice site $x$,
but it requires prohibitively large computational cost.
Instead, we may construct the all-to-all quark propagator,
which contains propagations from any lattice site to any site,
in an effective way~\cite{A2A}.
We determine low-lying eigenvalues $\{\lambda_k\}$ 
and corresponding eigenvectors $\{u_k\}$ of $D$,
and the contribution of these low-modes to the all-to-all quark propagator 
is calculated exactly as 
$S_{F,\rm low}(x,y)  
 = 
 \sum_{k=1}^{N_e}
 \frac{1}{\lambda_k} u_k(x) u_k^\dagger(y)$.
The number of the low-modes $N_e$ is set to 160 (240) on the smaller 
(larger) lattice. 
We confirm that 
low-energy observables, such as disconnected hadron correlators, 
are well dominated by the low-mode contribution with this choice of $N_e$.
The small contribution from the higher-modes
can be estimated stochastically by the so-called noise method, 
which is not computationally intensive.
This all-to-all propagator is used to calculate 
disconnected correlators of the flavor-singlet mesons and nucleon.
We refer to Refs.~\cite{FlvSngl:Nf3:JLQCD,StrCnt:Nf2:JLQCD:direct,PFF:Nf2:RG+Ovr:JLQCD}
for technical details.


\section{Flavor-singlet meson spectrum}

\begin{figure}[t]
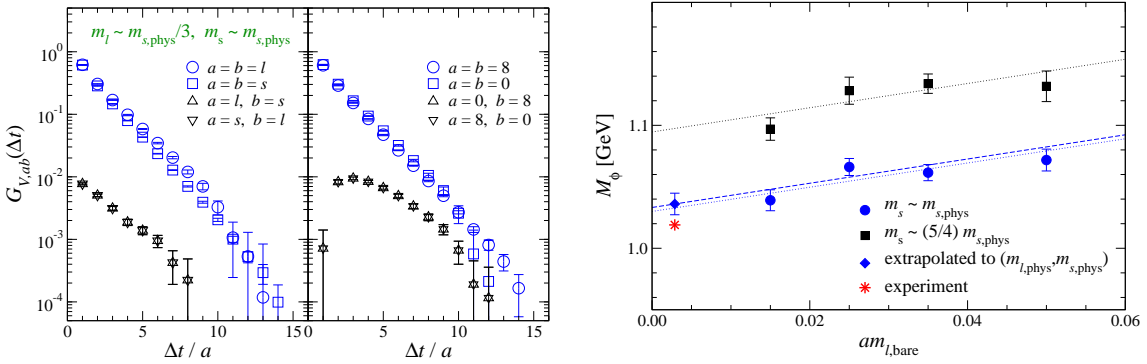

\begin{center}
   \includegraphics[angle=0,width=0.48\linewidth,clip]%
                   {msn_vv_mud1_ms6_smr11.eps}
   \hspace{4mm}
   \includegraphics[angle=0,width=0.48\linewidth,clip]%
                   {Mphi_vs_ml.eps}
   \vspace{-7mm}
   \caption{
      Left panel:
      vector meson correlators $G_{V,ab}$. 
      Right panel: 
      chiral extrapolation of $M_\phi$.
      The horizontal line represents the bare light quark mass 
      in lattice units, and $m_{s,\rm phys}$ is 0.081 in this normalization.
   }
   \label{fig:flv-sngl}
\vspace{-5mm}
\end{center}
\end{figure}

In the left panel of Fig.~\ref{fig:flv-sngl},
we plot vector meson correlators 
$G_{V, ab}(\Delta t)
 \!=\!(1/N_t)\sum_t \langle V_a(t+\Delta t) V_b^\dagger(t) \rangle$
of the light $(V_l\!=\! \sqrt{2} \bar{l} \gamma_k l)$ and 
strange vector mesons $(V_s\!=\! \bar{s}\gamma_k s)$ as well as 
those of the flavor-octet 
$(V_8\!=\! \sqrt{2/3}(\bar{l} \gamma_k l - \bar{s} \gamma_k s))$
and 
singlet mesons
$(V_0\!=\! (2\bar{l} \gamma_k l + \bar{s} \gamma_k s)/\sqrt{3})$.
The off-diagonal correlators $G_{V,\{ls,sl\}}$ of $V_l$ and $V_s$ 
are about two orders of magnitude smaller than 
the diagonal ones $G_{V,\{ll,ss\}}$, while 
there is no such large hierarchy between $V_0$ and $V_8$.
We can express $\omega$ and $\phi$ meson states
as eigenvectors of the $2 \!\times\! 2$ correlator matrices $G_{V,ab}$
($a,b \!\in\! \{l,s\}$ or $\{8,0\}$)
\bea
   \left\{
   \begin{array}{ll}
      \phi   \ = \ 1.00(1)\,V_s - 0.04(9)\,V_l
             \ = \ 0.84(5)\,V_8 - 0.55(7)\,V_0
      \\
      \omega \ = \ 0.04(9)\,V_s + 1.00(1)\,V_l
             \ = \ 0.55(7)\,V_8 + 0.84(5)\,V_0
   \end{array}
   \right..
   \label{eqn:sngl:omega_phi}
\eea
These observations imply an almost ideal mixing of the vector mesons,
which is expected from the experimental spectrum.
The right panel of Fig.~\ref{fig:flv-sngl} shows
a linear chiral extrapolation of the mass of the $\phi$ meson state
in (\ref{eqn:sngl:omega_phi}).
At the physical quark masses,
we obtain $M_\phi   = 1036(12)_{\rm stat}(-91)_{\rm sys}~\mbox{MeV}$
and $M_\omega = 854(24)_{\rm stat}(-92)_{\rm sys}~\mbox{MeV}$,
where the systematic error is estimated 
by adding higher order terms into the extrapolation form.

A similar analysis for the PS mesons suggests that 
$\eta$ and $\eta^\prime$ mesons are significant mixture
of the light ($P_l$) and strange mesons ($P_s$)
in contrast to the vector mesons: 
\bea
   \eta        = 0.96(1)\,P_l - 0.28(3)\,P_s,
   \hspace{3mm}
   \eta^\prime = 0.28(3)\,P_l - 0.96(1)\,P_s.
   \label{eqn:ps:mix}
\eea
A linear chiral extrapolation yields 
$M_\eta \! = \! 620(47)_{\rm stat}~\mbox{MeV}$ 
and $M_{\eta^\prime} = 789(128)_{\rm stat}~\mbox{MeV}$.
The obtained spectrum of the vector and pseudo-scalar mesons
is in good agreement with experiment.


\section{Nucleon strange quark content}

\begin{figure}[t]
\begin{center}
   \includegraphics[angle=0,width=0.48\linewidth,clip]{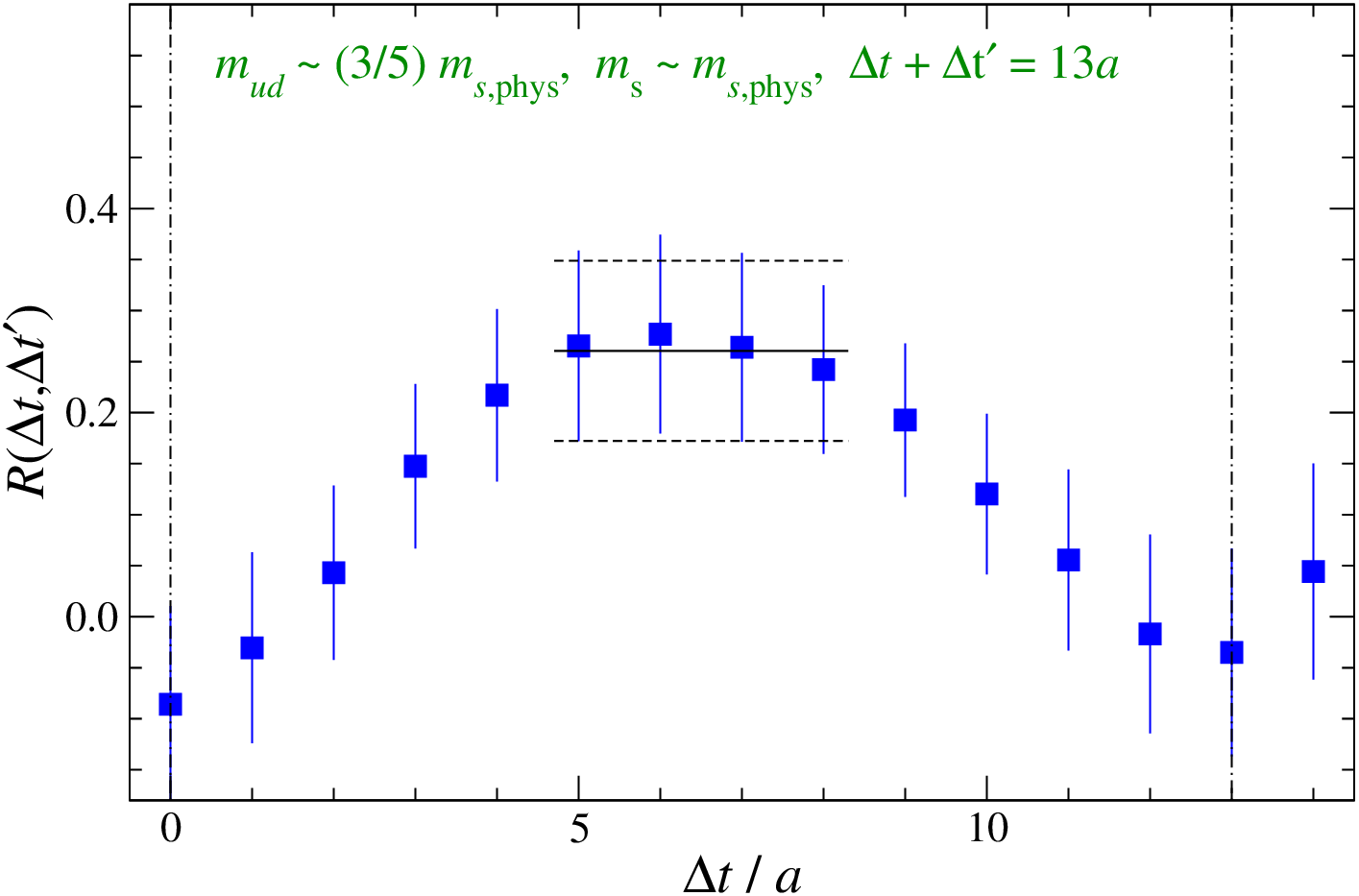}
   \hspace{4mm}
   \includegraphics[angle=0,width=0.48\linewidth,clip]{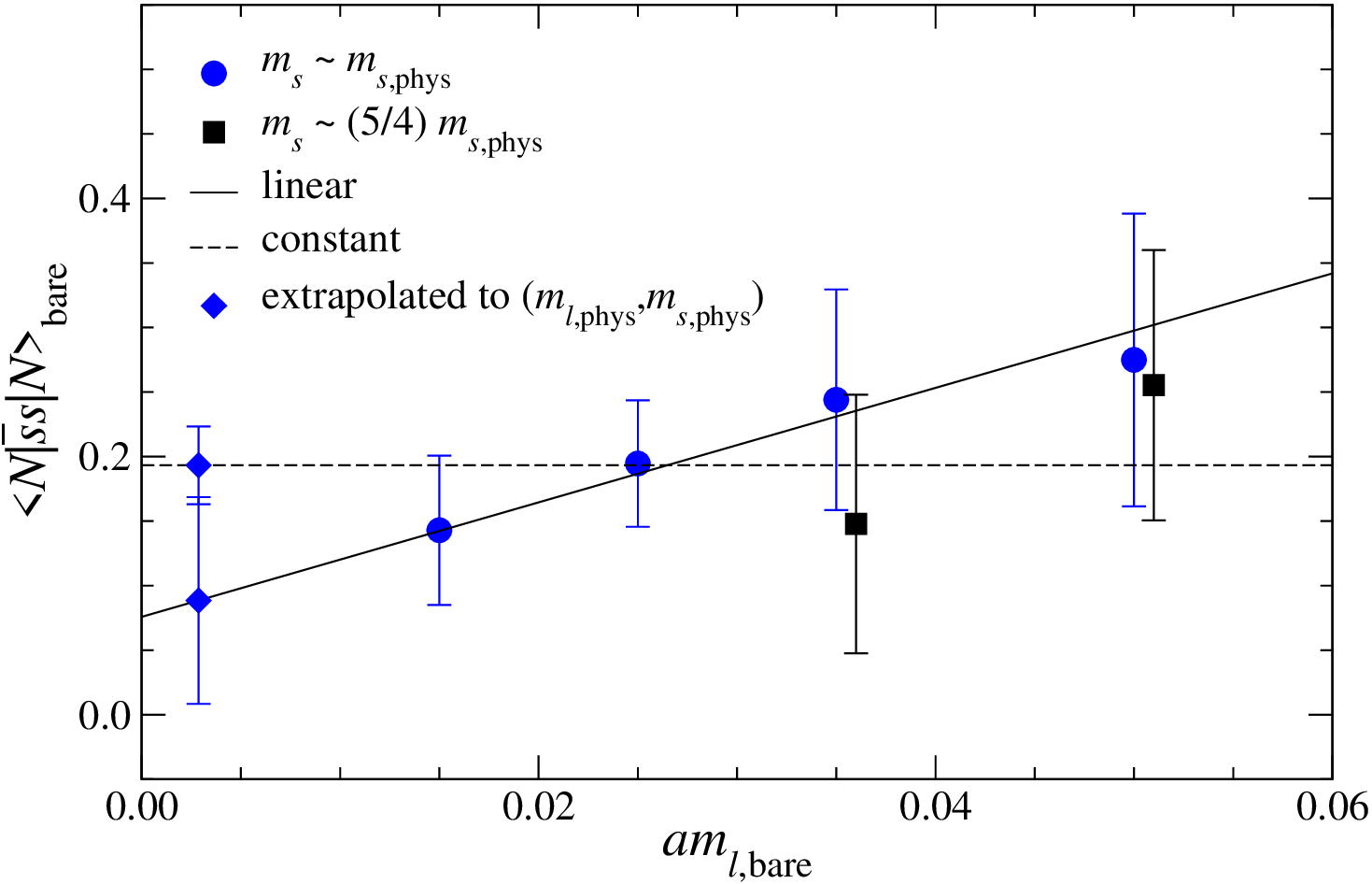}
   \vspace{-7mm}
   \caption{
      Left panel:
      ratio $R(\Dt,\Dtp)$ 
      with $\Dt\!+\!\Dtp\!=\!13a$ as a function of $\Dt$.
      Right panel:
      chiral extrapolation of $\NssNbare$.  
      Solid and dashed lines show linear and constant fits 
      at $m_s\!=\!m_{s,\rm phys}$.
   }
   \label{fig:str_cnt}
   \vspace{-5mm}
\end{center}
\end{figure}

We extract the bare value of the strange quark content $\NssNbare$ 
from the ratio of the three- and two-point functions 
\bea
   R(\Dt,\Dtp)
   & = &
   \frac{\sum_t \langle \calO_N(t+\Dt+\Dtp) | \{\bar{s}s(t+\Dt)\}_{\rm bare} | \overline{\calO}_N(t) \rangle}
        {\sum_t \langle \calO_N(t+\Dt+\Dtp) | \overline{\calO}_N(t) \rangle}
   \hspace{1mm} \xrightarrow[\Dt,\Dtp \to \infty]{} \hspace{1mm}
   \NssNbare.
   \label{eqn:str_cnt:ratio}
\eea
A clear non-zero signal shown in the left panel of Fig.~\ref{fig:str_cnt}
is obtained 
by improving our simulation set up \cite{StrCnt:Nf2:JLQCD:direct}: 
namely, we use the all-to-all propagator to precisely calculate 
the disconnected quark loop, 
and also use smeared nucleon source ($\bar{\calO}_N$) 
and sink operators ($\calO_N$) in order to 
reduce excited state contamination of the three- and two-point functions.

In our previous study in two-flavor QCD \cite{StrCnt:Nf2:JLQCD:direct},
we observe that 
the heavy baryon chiral expansion of $\NssNbare$ 
shows poor convergence at similar quark masses. 
We therefore employ a simple linear extrapolation to the physical point 
and also test a constant fit to estimate the systematic uncertainty.
Those fits are shown in the right panel of Fig.~\ref{fig:str_cnt}.
The extrapolated value 
is converted to a renormalization group invariant parameter
\begin{eqnarray}
  f_{T_s}
  =
  \frac{m_s \NssN}{M_N}
  =
  0.013(12)_{\rm stat}(16)_{\rm sys}.
  \label{Eqn:fts}
\end{eqnarray}
We note that chiral symmetry, which is preserved in our simulation,
is crucial to avoid a possibly large contamination 
due to the operator mixing with the light quark content $\NllN$
\cite{StrCnt:Nf2:JLQCD:direct}.
We also emphasize that 
(\ref{Eqn:fts}) is in good agreement with 
our previous estimate in two-flavor QCD \cite{StrCnt:Nf2:JLQCD:direct}
and our indirect determination through the Feynman-Hellmann theorem 
\cite{StrCnt:Nf2_3:JLQCD:spectrum}: 
all our studies consistently favor the small strange quark content
$f_{T_s}\!\approx\!0.01$\,--\,0.03.
 

\section{Conclusion}


In this article,
we demonstrated 
the feasibility of a quantitative study of 
the flavor-singlet spectrum and nucleon strange quark content
by using the all-to-all propagator.
It is interesting to extend this study to other observables,
such as the nucleon spin fraction carried by sea quarks.

We also note that
the pion scalar form factor has a significant contribution
from the disconnected diagram \cite{PFF:Nf2:RG+Ovr:JLQCD}.
Although disconnected contributions are generally (much)
smaller than connected ones,
they can not be naively ignored 
in future precise determinations of hadron observables.

\vspace{3mm} 

Numerical simulations are performed on Hitachi SR11000 and 
IBM System Blue Gene Solution 
at High Energy Accelerator Research Organization (KEK) 
under a support of its Large Scale Simulation Program (No.~09/10-09).
This work is supported in part by the Grant-in-Aid of the
Japanese Ministry of Education, Culture, Sports, Science and Technology
(No.~20105005 and 21684013).


\end{document}